\documentclass[pra,12pt,english]{article}
\usepackage{graphicx}

\begin{document}

\title{Weak Value and Weak Measurements}

 \maketitle \vskip 20 pt

The weak value of a variable $O$ is a description of an effective
interaction with that variable in the limit of weak coupling. For a
pre- and post-selected system described at time $t$ by the two-state
vector  $ \langle \Phi |~~|\Psi\rangle $ (see entry Two-State Vector
Formalism), the weak value is \cite{s-100}:
\begin{equation}\label{wv}
O_w \equiv { \langle{\Phi} \vert O \vert\Psi\rangle \over
\langle{\Phi}\vert{\Psi}\rangle } .
\end{equation}

Contrary to classical physics, variables in quantum mechanics might
not have definite values at a given time. In the complete
description of a usual (pre-selected) quantum system, the state
$|\Psi\rangle $ yields probabilities $p_i$ for various outcomes
$o_i$ of (an ideal) measurement of the variable $O$. Numerous
measurements on an ensemble of identical systems yield an average --
expectation value of $O$: $ ~\sum p_i o_i$. Since $  p_i~= ~|\langle
O=o_i~|\Psi\rangle|^2$, the expectation value  can be expressed as
$\langle \Psi|O|\Psi\rangle$. If the coupling to the measuring
device is very small, this expression is related directly to the
response of the measuring device, and the measurement does not
reveal the eigenvalues $o_i$ and their probabilities $p_i$ .
Specifically, $\langle \Psi|O|\Psi\rangle$ is  the shift of the
quantum state of the pointer variable of the measuring device,
which, otherwise, is not distorted significantly due to the
measurement interaction.

For pre- and post-selected quantum system, the response of the
measuring device or any other system coupled weakly to the variable
 $O$, is the shift of the quantum state by the weak value (\ref{wv}). The
 coupling can be modeled by the von Neumann measurement interaction
\begin{equation}
  \label{neumann}
 H = g(t) P O,
\end{equation}
where $g(t)$ defines the time of the interaction, $\int g(t) =1$,
and $P$ is conjugate to the pointer variable $Q$. The weakness of
the interaction is achieved by choosing the wave function of the
measuring device so that $P$ is small. Small value of $P$ requires
also a small uncertainty in $P$, and thus a large uncertainty of
 the pointer variable $Q$ in the initial state and consequently, a large
uncertainty in the measurement. Therefore, usually, we need a large
ensemble of identical pre- and post-selected quantum systems in
order to measure the weak value.

For rare post-selection, when $|\langle \Phi|\Psi\rangle|\ll 1$,
 the weak
value (\ref{wv}) might be far away from the range of the eigenvalues
of $O$, so it clearly has no statistical meaning as an ``average''
of $o_i$. If we model the initial state of the pointer by a Gaussian
$ \Psi^{MD}_{in} (Q) =(\Delta ^2 \pi )^{-1/4} e^{ -{{Q ^2} /{2\Delta
^2}}} $ with large $\Delta$ ensuring small $P$, the final state, to
a good approximation, is the shifted Gaussian $\Psi^{MD}_{fin} (Q) =
(\Delta^2 \pi )^{-1/4} e^{ -{{(Q - O_w)^2} /{2\Delta ^2}}}.$ The
standard  measurement procedure with weak coupling  reveals only the
real part of the weak value, which is, in general, a complex number.
Its imaginary part can be measured by observing the shift in $P$,
the conjugate to the pointer variable \cite{WV}.

The real part of the weak value is the outcome of the standard
measurement procedure at the limit of weak coupling. Unusually large
outcomes, such as spin 100 for a spin$-{1\over 2}$ particle
\cite{s-100}, appear from peculiar interference effect (sometimes
called Aharonov-Albert-Vaidman (AAV) effect) according to which, the
superposition of the pointer wave functions shifted by  small
amounts yields similar wave function shifted by a large amount. The
coefficients of the superposition are universal for a large class of
functions for which the Fourier transforms is well localized around
zero.

In the usual cases, the shift is much smaller than the spread
$\Delta$ of the initial state of the measurement pointer. But for
some variables, e.g. averages of variables of a large ensemble, for
very rare event in which all members of the ensemble happened to be
in the appropriate post-selected states, the shift is of the order,
and might be even larger than the spread of the quantum state of the
pointer \cite{AACV}. In such cases the weak value is obtained in a
single measurement which is not really ``weak''.

One can get an intuitive understanding of the AAV effect, noting
that the coupling of the weak measurement procedure  does not change
significantly the forward and the backward evolving quantum states.
Thus, during  the interaction, the measuring device ``feels'' both
forward and backward evolving quantum states. The tolerance of the
weak measurement procedure to the distortion due to the measurement
depends on the value of the scalar product  $ \langle \Phi
|\Psi\rangle $.

Since the quantum states remain effectively unchanged during the
measurement, several weak measurements can be performed one after
another and even simultaneously.  ``Weak-measurement elements of
reality'' \cite{WMER}, i.e., the weak values, provide self
consistent but sometimes very unusual picture for pre- and
post-selected quantum systems. Consider a three-box paradox in which
a single particle in three boxes is described by the two-state
vector
\begin{equation}
  \frac{1}{3}
\left(\left\langle A\right|+\left\langle B\right|-\left\langle
C\right|\right)~~~~\left(\left|A\right\rangle
 +\left|B\right\rangle
+\left|C\right\rangle \right) ,\label{3box}\end{equation} where
$\left|A\right\rangle $ is a quantum state of the particle located
in box $A$, etc. Then, there are the following weak-measurements
elements of reality regarding projections on various boxes:
$(\textbf{P}_A)_w=1$, $(\textbf{P}_B)_w=1$, $(\textbf{P}_C)_w=-1$.
Any weak coupling to the particle in box $A$ behaves as if there is
a particle there and the same for box $B$. Finally, a   weak
measuring device coupled to the particle in box $C$ is shifted by
the same value, but in the opposite direction. The coupling to the
projection onto all three boxes,
$\textbf{P}_{A,B,C}=\textbf{P}_A+\textbf{P}_B+\textbf{P}_C$
``feels'' one particle:
$(\textbf{P}_A+\textbf{P}_B+\textbf{P}_C)_w=(\textbf{P}_A)_w+
(\textbf{P}_B)_w+(\textbf{P}_C)_w=1$.

 There have been numerous experiments showing weak values
\cite{Ex1,Ex2,Ex3,Ex4}, mostly of photon polarization and the AAV
effect has been well confirmed. Unusual weak values were used for
explanation peculiar quantum phenomena, e.g., superluminal velocity
of tunneling particles \cite{tun1,tun2}. It was suggested that the
type of an amplification effect which takes place for unusually
large weak values might lead to practical applications.

\vskip .5cm Lev Vaidman\hfill\break School of Physics and Astronomy
\hfill\break Raymond and Beverly Sackler Faculty of Exact
Sciences\hfill\break Tel-Aviv University, Tel-Aviv 69978, Israel

\end{document}